# Study on the Data Storage Technology of Mini-Airborne Radar Based on Machine Learning

Haishan Tian[1,2], Qiong Yang[2], Huabing Wang[1a], Jingke Zhang[1]

1  School of Physics and Electronics, Hunan Normal University, Changsha 410081, People's Republic of China

2  State Key Laboratory of Complex Electromagnetic Environment Effects on Electronics and Information System（CEMEE）, Luoyang 471000, People's Republic of China.

**Keywords**: machine learning; data storage; file management; airborne radar.

## Summary

The data rate of airborne radar is much higher than the wireless data transfer rate in many detection applications, so the onboard data storage systems are usually used to store the radar data. Data storage systems with good seismic performance usually use NAND Flash as storage medium, and there is a widespread problem of long file management time, which seriously affects the data storage speed, especially under the limitation of platform miniaturization. To solve this problem, a data storage method based on machine learning is proposed for mini-airborne radar. The storage training model is established based on machine learning, and could process various kinds of radar data. The file management methods are classified and determined using the model, and then are applied to the storage of radar data. To verify the performance of the proposed method, a test was carried out on the data storage system of a mini-airborne radar. The experimental results show that the method based on machine learning can form various data storage methods adapted to different data rates and application scenarios. The ratio of the file management time to the actual data writing time is extremely low.

## 1. Introduction

The airborne radar has a high data rate, which can reach hundreds of Mbps or even Gbps [1,2,3] in many applications. Such a high data rate has far exceeded the current wireless data transmission rate between air and ground [4,5], so the on-board data storage systems are required by the airborne radar.

NAND Flash has many excellent characteristics, such as small size, light weight, low power consumption, high storage density, fast data access speed, and good seismic performance [6,7,8,9,10,11,12]. So, it is widely used in the aerospace and other high-performance data storage fields [13,14,15,16]. However, the data storage systems based on this medium generally have a problem that file management time is long, which reduces the storage speed [17,18,19,20,21]. Therefore, the study on the technology of the high-speed and massive data storage under the condition of miniaturization has important practical value for the application of airborne radar.

Scholars from all over the world have carried out many researches on file management optimization methods. The airborne storage usually uses a certain fixed method [22,23,24,25,26,27,28,29]. The typical ones are the pre-allocation or post-allocation methods, such as all clusters pre-allocation (ACPA) [29], file allocation table (FAT) pre-allocation and file directory table (FDT) quasi-allocation (FPFQA) [22], FAT post- and FDT post- allocation (FPFPA) [19] and other methods. The ACPA method pre-allocates all unused free clusters before one file is stored, and releases the unused space after the storage [29]. This method greatly reduces the updating frequency of the FAT and the FDT when the data in this file are written. It is not suitable for the application where the free storage space is much larger than one file, because in this case the contents of the FAT will be frequently pre-allocated and released. The FPFQA method pre-allocates all unused space in the FAT before data storage, and releases the unused space after data storage [22]. Therefore, there is no FAT updating during the data storage, and the updating frequency of the FDT is also significantly reduced. However, this method is only suitable for the application that all the data are stored using a certain fixed pattern, and is not suitable for the most storages. The FPFPA method writes the file data in the FAT and the FDT after all the 16 files' data has been written [19], which greatly reduces the updating frequency of the file areas, but it is not suitable for the storage of the multiple data with various rates. Therefore, these methods generally have the problem of limited application.

Therefore, the study on data storage methods based on machine learning is carried out for the mini-airborne radar. Through the machine learning of radar echo, signal processing results, motion sensor data, status data, working parameters and other data, the storage method and parameters for each application is determined and can be adjusted in real-time. Finally, the efficient storage of airborne radar data is realized.

## 2. Background

The structure of the storage's file system is shown in Fig.1, which mainly includes two mutual-backup FATs, one FDT, and the data region [22,30], which are respectively used to establish the cluster link relationship, describe file directory information, and store data. The FAT uses a 4B space to describe the link relationship of a cluster, and the FDT uses a 32B space to describe the information of a file. The cluster in the data region is the basic unit of file management and contains dozens of sectors [31,32]. The size of sector is 512B.

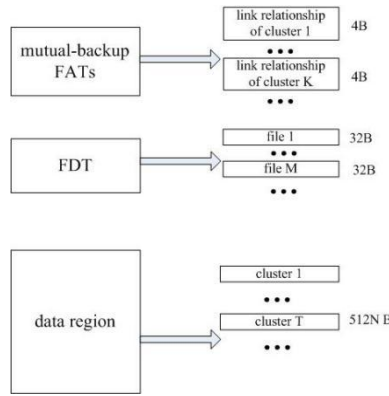

**Fig.1.** The structure of file management system in the data storage

The storage of one cluster of data includes operations of three areas, namely the FAT, the FDT and the data region. The data storage time is the sum of the data updating time of the three areas.

The updating time of the data region is

$$T_{DATA\_write} = (512 \times C \times T_w) + T_{w-res} \tag{1}$$

where C is the number of sectors in one cluster, generally dozens, $T_w$ is the average writing time of 1B data, and $T_{w-res}$ is the address jump response time. Generally, $T_{w-res}$ is a long time, which is related to the working mechanism of NAND flash, because the original data in the area must be erased firstly when writing data in the flash memory.

The two FATs only need to update 4B data respectively, but the sector with the size of 512B is the smallest unit of data reading and writing in the flash. So, the data of one sector need to be updated in FAT. The data updating time of the two FATs is

$$T_{FAT\_write} = 2(512 \times T_w + T_{w-res}) = 1024 T_w + 2T_{w-res} \tag{2}$$

Because $T_{w-res}$ is generally dozens of or even hundreds of times the data writing time of a sector, the real data writing time of the FAT can be ignored. So, $T_{FAT\_write}$ approximately equal to $2T_{w-res}$.

Similarly, the FDT that only needs to update the 32B data also writes the data of one sector. The data updating time of the FDT is

$$T_{FDT\_write} = 512 \times T_w + T_{w-res} \approx T_{w-res} \tag{3}$$

Therefore, the updating time of the three areas in the data storage of one cluster is shown in Table 1.

Table 1 Data updating time of each area in the traditional FAT file system

| | data region | FAT | FDT |
|---|---|---|---|
| data writing time per cluster | $512 C T_w + T_{w-res}$ | $2T_{w-res}$ | $T_{w-res}$ |

It can be seen from the Table 1 that response time of address jump brought by file management is $4T_{w-res}$, which even far exceeds the actual data writing time. The ratio of file management time to the actual data writing time $\mu = 4T_{w-res}/(512 \bullet C \bullet T_w)$ can reach dozens or even hundreds, which will significantly reduce the data storage speed.

## 3. Airborne radar data storage based on machine learning

To reduce the file management time in the data storage of mini-airborne radar, a storage method based on machine learning is proposed.

3.1 Data preprocessing

Machine learning is generally divided into supervised learning and unsupervised learning based on whether the processed data is artificially labeled. Supervised learning uses labeled data as learning data, which usually works well and requires less hardware computing and cache resources. Therefore, the data is preprocessed before machine learning to generate the labeled data.

Airborne radars usually generate echoes, signal processing results, sensor data, navigation data, status data, parameters, and other data. In addition to the above-mentioned data to be stored, data storage performance is also related to hardware resources such as the system's calculations, caches, storage medium, and interfaces. The impact of radar data and hardware resources on storage performance is studied, and the labeled data $X = [x_1, x_2, ..., x_m]$ is made in the data preprocessing.

According to the file management method, the labeled data $X$ should include 8 types of data: file size, packet size, data rate, data type, interface type and data transmission rate, computing resources, cache resources and storage resources.

- File size affects the updating frequency of the FDT;
- The packet size determines the size of the continuous writing data in the data region;
- Storage of different data will form data switching. This operation and the data rate affect the updating frequency of the FAT and the FDT;
- The interface type and data transmission rate are related to the buffer time during storage;
- Computing resources are related to the efficiency of operations such as calculation of storage space logical address, file naming, and state machine execution;
- The cache resources determine the data waiting time when storing various data, which is related to the querying and updating strategy of the FAT and the FDT;
- The type and array pattern of storage medium are related to the transmission layer algorithm in the FAT, FDT and data region.

Therefore, data preprocessing is first performed before data storage, and the data to be stored and hardware resources are processed to the labeled data.

3.2 Data storage of the airborne radar based on machine learning

A method based on machine learning is proposed to achieve efficient storage of airborne radar data in this study. The process of the method is shown in Fig.2, which includes an input layer, multiple hidden layers, and an output layer. The statistical features in the data are extracted through layer-by-layer training and nonlinear transformation of supervised data. Then the iterative calculations from right to left through supervised learning is established to complete the overall network tuning. So, the mapping relationship from the low-level signal to the high-level data storage method is established, and the network initialization is completed.

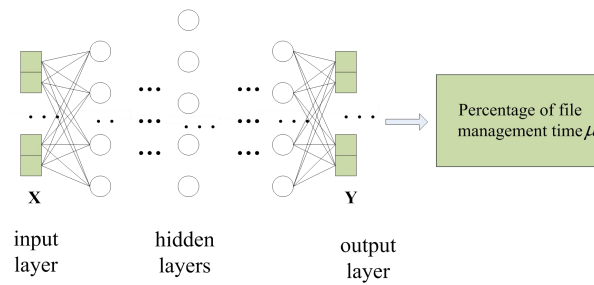

**Fig.2.** The flow chart of airborne radar data storage method based on machine learning

The signal of input layer in Fig.2 is $X = [x_1, x_2, ..., x_8]$, which contains 8 types of the preprocessed data. Each represents a kind of data which is a vector signal.

The activation function of the hidden layers determines the form of the non-linear feature mapping, and $f^l, l \in N_l$ is used to represent the activation function vector of each layer, where $N_l$ is the number of hidden layers in Fig.2. $W^l$ and $b^l$ represent the weight vector and bias vector of the layer $l$ respectively. The specific steps are introduced as follows.

The method of layer-by-layer training is adopt to realize machine learning: input training data $X$ from the input layer, train the network parameters of the first hidden layer, and activate the parameters $X_1$ as the output of the first hidden layer, which is equal to

$$X_1 = f^1(W^1 X + b^1) \tag{4}$$

Then use the output of the first hidden layer as the input of the second hidden layer, and train the output of the second hidden layer, $X_2$, in the same way, which is equal to

$$\begin{aligned} X_2 &= f^2(W^2 X + b^2) \\ &= f^2(W^2(f^1(W^1 X + b^1)) + b^2) \end{aligned} \tag{5}$$

Repeat this step until the output of all hidden layers is trained. The output of the last hidden layer is

$$X_{l-1} = f^{l-1}(W^{l-1}(f^{l-2}(W^{l-3} \cdots f^1(W^1 X + b^1)) + b^{l-2}) + b^{l-1}) \tag{6}$$

The output of the last hidden layer is used as the input of the output layer, and the final data output is

$$Y = f^l(W^l(f^{l-1}(W^{l-2} \cdots f^1(W^1 X + b^1)) + b^{l-1}) + b^l) \tag{7}$$

Among them, Y contains 5 kinds of data, which are FAT query method, FAT update method, FDT query method, FDT update method, and data region write method. The network parameters of all hidden layers and output layer, which is regarded as the parameters of a network learning model, would be integrated and all initialized.

The cost function of the entire network, $E(X,Y)$, is solved iteratively through optimization theory. $E(X,Y)$ is the ratio of the file management time to the actual data writing time. All network parameters are gradually fine-tuned, and finally the optimized parameters are obtained, which are used for subsequent classification or prediction.

After obtaining the file management method set Y, the airborne radar data storage is executed using the method in Y.

## 4. Experimental test

The proposed data storage method based on machine learning was tested on a storage system of the airborne synthetic aperture radar, as shown in Fig.3. The radar is installed on the plane's diagonal rod, as shown in the yellow box of Fig.3(a), and the red box is the storage system. Fig.3(b) is the internal diagram of the storage system. The CF card is the storage medium, the FPGA is the processor that controls the operations of the data storage, and the DDR3 SDRAM is the data cache. The data to be stored include the radar echo output by ADC, signal processing results, GPS data, sensor data, USB data, system status data, and working parameters. These data are received or acquired by the FPGA.

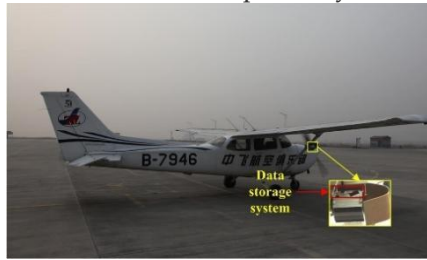

(a) The airborne synthetic aperture radar

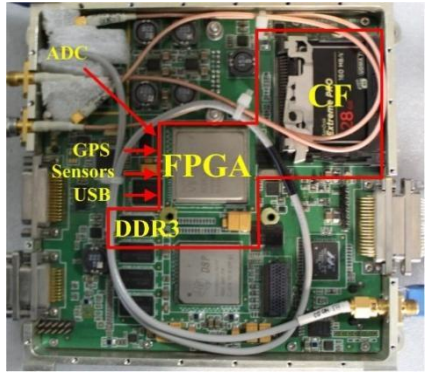
(b) The data storage system of airborne radar
**Fig.3.** A mini-airborne synthetic aperture radar and data storage system

3.1 Training data set

The training data set consists of the real data generated or received by the data storage system shown in Fig.3(b), which contains 7 types of data. The data generation method, the size and interval of the data rate, and the size and interval of the continuously written data packets are shown in Table 2.

Table 2 Training data set of airborne storage based on machine learning

| data type | data rate /its interval | packet size /its interval | generation method | training data number |
|---|---|---|---|---|
| radar echo | 1MB/s-120MB/s /interval 2MB/s | 16kB-256kB /interval 16kB | adjust sampling rate of the ADC | 960 |
| processing results | 1MB/s-32MB/s /interval 1MB/s | 16kB-256kB /interval 16kB | generate in FPGA | 512 |
| status data | 16kB/s-1MB/s /interval 16kB/s | 1kB-32kB /interval 2kB | generate in FPGA | 1024 |
| GPS data | 2kB/s-10kB/s /interval 2kB/s | 1kB-32kB /interval 1kB | adjust data rate of GPS receiver | 160 |
| sensor data | 2kB/s-10kB/s /interval 2kB/s | 1kB-32kB /interval 1kB | adjust data rate of the sensor | 160 |
| working parameters | 16kB/s-256kB/s /interval 16kB/s | 1kB-32kB /interval 1kB | generate in FPGA | 512 |
| USB data | 128kB/s-16MB/s /interval 128kB/s | 16kB-128kB /interval 16kB | adjust data rate of the USB | 1024 |

The training data set in Table 2 covers the radar data to be stored and their parameters, such as data type, data rate, data packet size and so on. The data set is firstly preprocessed on the storage system shown in Fig.3(b) to generate labeled data, and then the data storage based on machine learning is carried out to classify and determine file management methods. Finally, a data storage model based on machine learning is formed in the airborne data storage system.

3.2 The data storage test of airborne radar

To verify the performance of the proposed data storage method, tests of four methods, which includes the storage method based on machine learning, the original FAT file system, the ACPA method, the FPFQA method, and the FPFPA method, were performed on the system shown in Fig.3(b). Several common different types of data or their combinations were used in the test. The data types and results are shown in Table 3. The data writing rate in the test is the highest data writing rate of the CF card, 160MB/s, and the size of the test data is 2GB. The value of $\mu$ in Table 3 is equal to the proportion of file management time to the actual data writing time, which is used as a parameter to measure the performance of four methods. This is because the time cost in data storage consists of two parts: the actual data writing time and the file management time. The actual data writing time of each method is fixed and the same, and the performance difference is determined by the file management time.

Table 3 Test results of the data storage system in a mini-airborne synthetic aperture radar

| storage data and data rate | original FAT $\mu$ | ACPA $\mu$ | FPFQA $\mu$ | FPFPA $\mu$ | machine learning $\mu$ |
|---|---|---|---|---|---|
| radar echo,16MB/s | 102.21 | 0.32 | $6\times10^{-4}$ | 0.026 | 0.017 |
| radar echo,80MB/s | 102.52 | 0.33 | $5\times10^{-4}$ | 0.016 | 0.013 |
| radar echo,16MB/s processing results,1MB/s | 102.15 | 0.51 | $9\times10^{-4}$ | 0.050 | 0.025 |
| radar echo,80MB/s processing results,5MB/s | 102.16 | 0.50 | $7\times10^{-4}$ | 0.038 | 0.021 |
| other 5 types of data, each data rate lower than 1MB/s, random storage | 102.78 | 26.06 | / | 20 | 0.11 |
| radar echo,80MB/s; other 5 types of data, each data rate lower than 1MB/s, random storage | 102.90 | 26.71 | / | 20.5 | 0.032 |

Remarks: '/'means this method is not applicable to the storage for this type of data.

According to the results in Table 3, the following conclusions can be drawn.

(1) The file management time of the original FAT file system is too long, and the ratio to the actual data writing time, $\mu$, is greater than 102. For the storage medium with the maximum data writing rate 160MB/s, the maximum data storage speed is only $160/(\mu+1)$ $MB/s \approx 1.55 MB/s$. So, the FAT file system is not suitable for airborne radar data storage;

(2) The $\mu$ of the ACPA method is dozens of or even hundreds of that of the data storage method based on machine learning;

(3) For the storage of fixed data such as radar echo and signal processing results, the values of $\mu$ in the FPFQA method, machine learning-based method and FPFPA method are all extremely low. Among them, the value of $\mu$ in the FPFQA method is the lowest. The $\mu$ in the machine learning method is lower than that in the FPFPA method, and the maximum is 0.025. The file management time of the three methods has little effect on this type of data storage.

(4) When low-rate data and the high-rate data are mixed and stored, the FPFQA method cannot be applied. This method is only applicable to the application where all data are stored in a predetermined way. The $\mu$ in the machine learning method for hybrid data storage is still low, and its $\mu$ is only a few tenths of that in the FPFPA method.

Therefore, the proposed machine learning-based storage method for the mini-airborne radar data has the strongest applicability. It is not only suitable for storage of the data with high data rate, but also for hybrid storage of the data with various data rates. What's more, its values of $\mu$ are low for all types of data storage, and have little impact on the data storage speed. So, the data storage method based on machine learning proposed in this study is suitable for data storage of mini-airborne radar.

# 5. Conclusion

Airborne radar data storage has the problem of long file management time, especially in the case that the mini-aircraft systems have limited computing and cache resources. The current methods are mainly pre-allocation or post-allocation ways, which are only suitable for one or several types of data storage, and their universality is not very well. Therefore, this study proposes a data storage method based on machine learning, which collects, processes, and forms a data training set to be used to establish a machine learning training model, and then processes the airborne radar data using the model, finally classifies and determines the applicable file management method. To verify the performance of the proposed method, an experimental test was carried out on the data storage system of a mini-airborne radar. Experimental results show that the file management time of the proposed data storage method based on machine learning is much shorter than the actual data writing time, and it is suitable for the storage of various types of data.

**Acknowledgments**
This research was funded by National Natural Science Foundation of China grant number 61801495 and National Natural Science Foundation of China grant number 62001486.

# References

1. H.L. Lu, Y.N. Li and H. Li, et al.: "Ship Detection by an Airborne Passive Interferometric Microwave Sensor (PIMS)," IEEE Geoscience and Remote Sensing, 58 (2020) 2682 (DOI: 10.1109/TGRS.2019.2953355).
2. E. Taghavi, D. Song, R. and Tharmarasa, et al.: "Geo-registration and Geo-location Using Two Airborne Video Sensors," IEEE Transactions on Aerospace and Electronic Systems, 56 (2020) 2910 (DOI: 10.1109/TAES.2020.2995439).
3. N. Duda, S. Ripperger and F. Mayer, et al.: "Low-Weight Noninvasive Heart Beat Detector for Small Airborne Vertebrates," IEEE Sensors Letters, 4 (2020) 2475 (DOI: 10.1109/LSENS.2020.2971769).
4. W.T. Chen and R.R. Mansour: "Miniature Gas Sensor and Sensor Array with Single- and Dual-Mode RF Dielectric Resonators," IEEE Transactions on Microwave Theory and Techniques, 66 (2018) 3697 (DOI: 10.1109/TMTT.2018.2854551).
5. Y. Ye and D. Pui: "Airborne Particle Detection Based on a Spin Hall Effect Measurement," IEEE Sensors Letters, 3 (2019) 2475 (DOI: 10.1109/LSENS.2019.2893123).

6. G. Lim, K.S. Park, N.C. Park, et al.: "Analysis of the Influence of the Ramp and Disk Dynamics on the HDI Response of 2.5-in Hard Disk Drive to a Shock," IEEE Transactions on Magnetics, 53 (2017) 3697 (DOI: 10.1109/TMAG.2016.2620182).
7. R.M. Brockie, T. Ngo; J.P. Chen, et al.: "ATI Considerations in Helium-Filled Hard Disk Drives," IEEE Transactions on Magnetics, 55 (2018) 2225 (DOI: 10.1109/TMAG.2018.2876648).
8. F. Wang, Y. Feng, X.P. Zhan, et al.: "Implementation of Data Search in Multi-Level NAND Flash Memory by Complementary Storage Scheme," IEEE Electron Device Letters, 41 (2020) 2280 DOI: 10.1109/LED.2020.3004989).
9. S. Kim, H. Kim, C. Woo, et al.: "Separation of Lateral Migration Components by Hole During the Short-Term Retention Operation in 3-D NAND Flash Memories," IEEE Transactions on Electron Devices, 67 (2020) 2645 (DOI: 10.1109/TED.2020.2989734).
10. J.Y. Park, D.H. Yun, S.Y. Kim, et al.: "Suppression of Self-Heating Effects in 3-D V-NAND Flash Memory Using a Plugged Pillar-Shaped Heat Sink," IEEE Electron Device Letters, 40 (2018) 212 (10.1109/LED.2018.2889037).
11. W. Hou, L. Jin, X.L. Jia, et al.: "Investigation of Program Noise in Charge Trap Based 3D NAND Flash Memory," IEEE Electron Device Letters, 41 (2019) 30 (DOI: 10.1109/LED.2019.2954621).
12. H.C. Hong, C.K. Yang, et al.: "Interleaved Write Scheme for Improving Sequential Write Throughput of Multi-Chip MLC NAND Flash Memory Systems," IEEE Transactions on Circuits and Systems I: Regular Papers, 67 (2020) 4946 (DOI:10.1109/TCSI. 2020.3015981).
13. Y.C. Kong, M. Zhang, X.P. Zhan, et al.: "Retention Correlated Read Disturb Errors in 3-D Charge Trap NAND Flash Memory: Observations, Analysis, and Solutions," IEEE Transactions on Computer-Aided Design of Integrated Circuits and Systems, 39 (2020) 4042 (DOI: 10.1109/TCAD.2020.3025514).
14. K. Lee, H. Shin, et al.: "Investigation of Retention Characteristics for Trap-Assisted Tunneling Mechanism in Sub 20-nm NAND Flash Memory," IEEE Transactions on Device and Materials Reliability, 17 (2017) 758 (DOI: 10.1109/TDMR.2017.2772046).
15. G.T. Raúl, et al.: "BenchBox: A User-Driven Benchmarking Framework for Fat-Client Storage Systems," IEEE Transactions on Parallel and Distributed Systems, 29(2018)2191 (DOI: 10.1109/ TPDS.2018.2819657).
16. L.P. Chang, C.H. Cheng, S.T. Chang, et al.: "Current-Aware Flash Scheduling for Current Capping in Solid State Disks," IEEE Transactions on Computer-Aided Design of Integrated Circuits and Systems, 39 (2018) 321 (DOI: 10.1109/TCAD.2018.2887046).
17. K. Han, H. Kim, D. Shin, et al.: "WAL-SSD: Address Remapping-Based Write-Ahead-Logging Solid-State Disks," IEEE Transactions on Computers, 69 (2019) 260 (DOI: 10.1109 /TC.2019.2947897).
18. L. Zuolo, C. Zambelli, A. Marelli, et al.: "LDPC Soft Decoding with Improved Performance in 1X-2X MLC and TLC NAND Flash-Based Solid State Drives," IEEE Transactions on Emerging Topics in Computing, 7 (2017) 507 (DOI: 10.1109/TETC.2017.2688079).
19. H.S. Tian, et al.: "Study on an on-board data storage system for FMCW SAR," Journal of Applied Remote Sensing, 10(2016)11.
20. D. Min, D. Park, J. Ahn, et al.: "Amoeba: An autonomous backup and recovery SSD for ransomware attack defense," IEEE Computer Architecture Letters. 17 (2018) 245. (doi: 10.1109/LCA.2018.2883431)
21. D. Huang, D. Han, J. Wang;, et al.: "Achieving Load Balance for Parallel Data Access on Distributed File Systems," IEEE Transactions on Computers, 67 (2017) 388 (DOI: 10.1109/TC.2017. 2749229).
22. H.S. Tian, F.F. Ju, H.S. Nie, et al.: "A new technology for real-time file system of high-speed storage system in airborne sensors," IEICE Electronics Express, 18 (2021).
23. S.Z. Wu, B. Mao, H. Jiang, et al.: "PFP: Improving the Reliability of Deduplication-based Storage Systems with Per-File Parity," IEEE Transactions on Parallel and Distributed Systems, 30 (2019) 2117 (DOI: 10.1109/TPDS.2019.2898942).
24. Y.X. Chen, C. Li, M. Lv, et al.: "Explicit Data Correlations-Directed Metadata Prefetching Method in Distributed File Systems," IEEE Transactions on Parallel and Distributed Systems, 30 (2019) 2692 (DOI: 10.1109/TPDS.2019.2921760).
25. J. Zhou, Y. Chen, W.P. Wang, et al.: "A Highly Reliable Metadata Service for Large-Scale Distributed File Systems," IEEE Transactions on Parallel and Distributed Systems, 31 (2019) 374 (DOI:10.1109/TPDS.2019.2937492).
26. J. Wang, D.Z. Han, J.Y. Zhang , et al.: "G-SD: Achieving Fast Reverse Lookup using Scalable Declustering Layout in Large-Scale File Systems," IEEE Transactions on Cloud Computing, 6 (2016) 1071 (DOI:10.1109/TCC.2016.2586050).
27. M. Sookhak, F.R. Yu, A.Y. Zomaya, et al.: "Auditing Big Data Storage in Cloud Computing Using Divide and Conquer Tables," IEEE Transactions on Parallel and Distributed Systems , 29 (2017) 999 (DOI: 10.1109/TPDS.2017.2784423).
28. Z.Z. Li, H.Y. Shen, et al.: "Measuring Scale-Up and Scale-Out Hadoop with Remote and Local File Systems and Selecting the Best Platform," IEEE Transactions on Parallel and Distributed Systems, 28 (2017) 3201 (DOI: 10.1109/TPDS.2017.2712635).
29. H. Park, et al.: "New techniques for real-time FAT file system in mobile multimedia devices," IEEE Transaction on Consumer and Electronic. 52, (2006)9 (DOI: 10.1109/TCE.2006.1605017).
30. S. Karki, B. Nguyen, J. Feener, et al.: "Enforcing End-to-End I/O Policies for Scientific Workflows Using Software-Defined Storage Resource Enclaves," IEEE Transactions on Multi-Scale Computing Systems, 4 (2018) 662 (DOI: 10.1109/TMSCS. 2018.2879096).
31. A. Boubriak, A. Cooper, C. Hossack, et al.: "SlimFS: A Thin and Unobtrusive File System for Embedded Systems and Consumer Products," IEEE Transactions on Consumer Electronics, 64 (2018) 334 (DOI: 10.1109/TCE.2018.2867826).
32. H. Sim, A. Khan, S.S. Vazhkudai, et al.: "An Integrated Indexing and Search Service for Distributed File Systems," IEEE Transactions on Parallel and Distributed Systems, 31 (2020) 2375 (DOI: 10.1109 /TPDS.2020.2990656).